\documentclass[prx,superscriptaddress,amsmath,amssymb,twocolumn]{revtex4}
\usepackage{mathrsfs}
\usepackage{graphicx}
\usepackage{dcolumn}
\usepackage[utopia]{mathdesign}
\let\mathbb\undefined  
\usepackage{bbold}     

\begin{document}

\title{On-chip superconducting microwave circulator from synthetic rotation}

\author{Joseph Kerckhoff}
\email{jakerckhoff@hrl.com}
\altaffiliation{Current address: HRL Laboratories, LLC, Malibu, CA 90265, USA}
\address{JILA, University of Colorado, Boulder, Colorado 80309, USA}
\author{Kevin Lalumi\`{e}re}
\address{D\'{e}partement de Physique, Universit\'{e} de Sherbrooke, Sherbrooke, Qu\'{e}bec, Canada J1K 2R1}
\author{Benjamin J. Chapman}
\address{JILA, University of Colorado, Boulder, Colorado 80309, USA}
\author{Alexandre Blais}
\address{D\'{e}partement de Physique, Universit\'{e} de Sherbrooke, Sherbrooke, Qu\'{e}bec, Canada J1K 2R1}
\address{Canadian Institute for Advanced Research, Toronto, Canada}
\author{K. W. Lehnert}
\address{JILA, University of Colorado, Boulder, Colorado 80309, USA}
\affiliation{National Institute of Standards and Technology, Boulder, Colorado 80305, USA}

\date{\today}

\begin{abstract}
We analyze the design of a potential replacement technology for the commercial ferrite circulators that are ubiquitous in contemporary quantum superconducting microwave experiments.  The lossless, lumped element design is capable of being integrated on chip with other superconducting microwave devices, thus circumventing the many performance-limiting aspects of ferrite circulators.  The design is based on the dynamic modulation of DC superconducting microwave quantum interference devices (SQUIDs) that function as nearly linear, tunable inductors.  The connection to familiar ferrite-based circulators is a simple frame boost in the internal dynamics' equation of motion.  In addition to the general, schematic analysis, we also give an overview of many considerations necessary to achieve a practical design with a tunable center frequency in the 4-8 GHz frequency band, a bandwidth of 240 MHz, reflections at the -20 dB level, and a maximum signal power of approximately order 100 microwave photons per inverse bandwidth.
\end{abstract}

\maketitle

\section{Introduction}

With the advent of quantum information processing with superconducting circuits~\cite{Devo13}, the ability to route microwave signals without loss or added noise has become critically important. In particular, the operation of a digital superconducting quantum computer will require numerous analog functions, such as signal amplification, feedback, and transduction. Moreover, it is likely that a future quantum information processor will employ coherent exchange of microwave fields among various modular components. Preserving quantum information in these propagating microwave fields demands nearly losses and noiseless components.

Although most passive microwave components can be readily fabricated with low-loss superconducting metals and integrated with other superconducting circuits, this is not true for non-reciprocal components such as isolators, circulators, and gyrators. Existing technology uses ferrimagnetic materials in intense ($\sim$0.1~T) magnetic fields to create the non-reciprocal behavior required to ensure the one-way flow of information within the network~\cite{Fay65, Alle56,Auld59}. Such magnetic devices would be quite difficult to integrate with superconducting circuits, which are disrupted by magnetic fields of 0.1~mT or less. Currently, the non-reciprocal elements in quantum information processing networks are commercially available devices connected to the rest of the network using meter length coaxial cables. Even if this cumbersome arrangement were tolerable, the loss associated with the transition from planar circuits to coaxial cables and that associated with the ferrimagnetic elements themselves is unacceptably large~\cite{Mall11,Murc12,Narl14,Roch14}.

Instead of achieving non-reciprocity through the use of magnetic materials, one can instead use time-dependent reactive elements.  Long known as a general method for creating non-reciprocal devices~\cite{Ande65,Ande66,Este14}, time varying reactances have not had much technological impact because ferrite elements~\cite{Fay65} or active transistor~\cite{Tana65} devices provide a less complex source of non-reciprocity for conventional electronics.  But with the emergence of superconducting circuit based quantum information processing, the idea of creating non-reciprocal response through time varying reactances has returned to prominence~\cite{Kama11,Fang12,Ranz14,Abdo14} because of the problems with ferrite devices and the practical absence of transistor technologies with quantum-limited noise performance~\cite{Cler10}.  

Conveniently, a standard superconducting circuit element, the Josephson junction, can be operated as a time-variable reactance.  Although the inductance of a Josephson junction is intrinsically non-linear, the inductance experienced by a small electrical signal is effectively linear, lossless, and may be varied by also applying a larger ``pump'' current through the junction~\cite{Yurk89}.  A new type of superconducting non-reciprocal device exploits this effect, in which several oscillatory pump tones modulate the inductance and therefore the frequency response of several resonant circuits in a cyclic manner~\cite{Kama11,Ranz14,Abdo14}.  Other novel approaches to nonreciprocal circuits for superconducting networks that don't rely on time-variable reactances are also active areas of research~\cite{Koch10,Viol14}.  

In any scheme based on time-variable reactances, it is important that the pump tones themselves do not interfere with the operation of other devices in the network~\cite{Abdo14}.  Furthermore the non-reciprocal device should not modulate the incoming signal to create sidebands that leave the device. And ideally, large signals should be processed linearly, despite the components' fundamental nonlinearities.  To these ends, we introduce another concept for a non-reciprocal element, a four-port circulator that operates by only modulating the coupling rate of four itinerant microwave modes to two resonant circuits in a cyclic manner.  This four-port circulator can also be wired as a two-port gyrator~\cite{Tell48}, and although not explored here, can thus also be wired as a three-port circulator~\cite{Fay65,Viol14,Carl55}.  The symmetry of the circuit ensures that sidebands generated by the dynamic coupling are completely ``erased'' as they leave the device. In addition, the modulating pump tones do not co-propagate with the signal tones and oscillate at a frequency at least a factor of ten less than the signal frequency, and can therefore be easily filtered out of the signal path between devices operating at the signal frequency. Finally, the Josephson junction elements are arranged into series arrays of junctions~\cite{Cast08}, an arrangement that retains the variable inductance but dilutes the non-linearity to approximate a linear time-variable inductor.

In this article, we first analyze the general equations of motion of a four port circulator created through time-dependent coupling between the ports and the internal resonant modes. We show that this concept is closely analogous to four-port ferrite circulators (where our time-varying reactances create a synthetic magnetic field \cite{Fang12}) and we use a waveguide ferrite circulator \cite{Alle56} as a pedagogical touchstone. We then introduce and analyze a circuit that realizes the desired circulator equations of motion. This non-reciprocal circuit uses a modular design comprising four identical subcircuits. These subcircuits are themselves composed of four tunable inductors arranged in a Wheatstone bridge configuration~\cite{Ande66,Berg10}, where the tunable inductors are created from flux tunable SQUID arrays. In the final section, we analyze SQUID arrays and how they form flux tunable inductors.

\section{Phenomenological description}\label{sec:Phenomenon}

\begin{figure*}
\includegraphics[width=.75\textwidth]{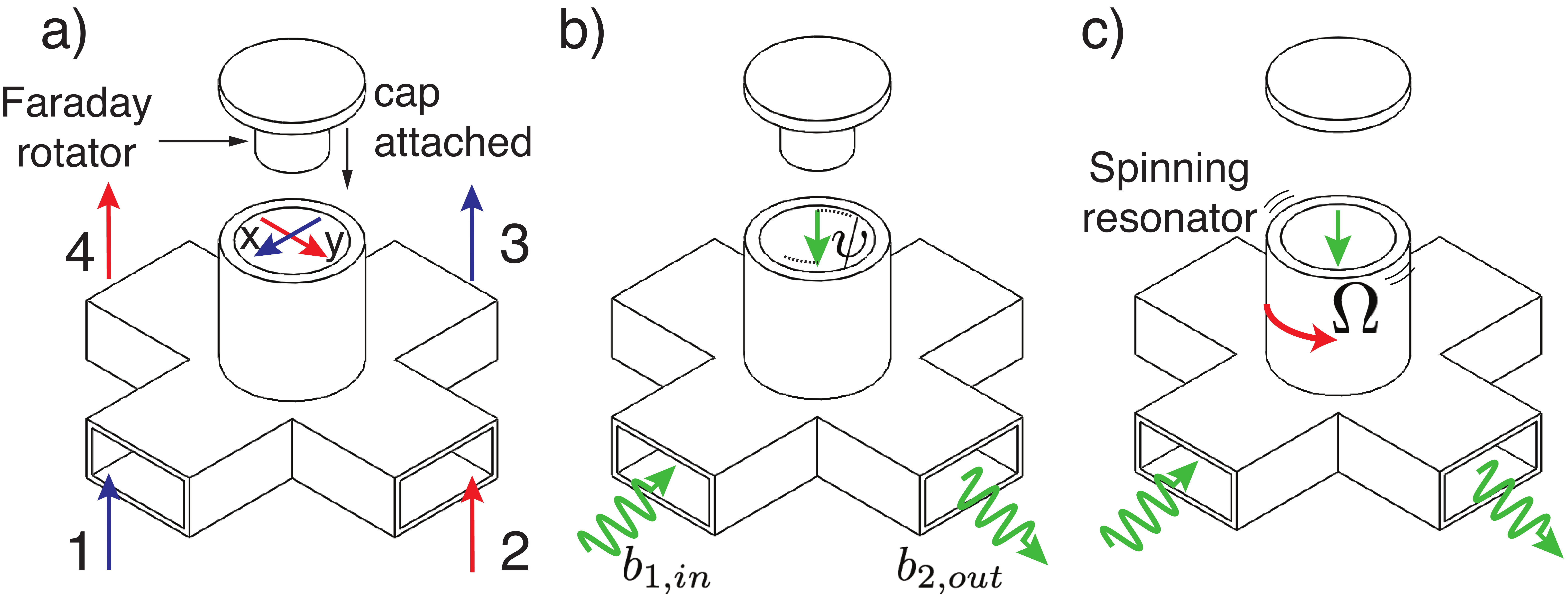}\\
\caption{a) Schematic of a four-port ferrite turnstile circulator \cite{Alle56, Auld59}.  Arrows represent the electric field polarization of the four waveguide and two resonator modes.  The electric fields of ports 1 \& 3 couple directly to resonator mode x (blue arrows), and the electric fields of ports 2 \& 4 couple directly to resonator mode y (red arrows).  b) Depiction of the circulator's steady state operation: a traveling wave signal incident through waveguide port 1 induces a steady state response in the resonator modes that is rotated by angle $\psi$ relative to the $x$-mode.  When $\psi=\pi/4$, all of the incident power is emitted out port 2.  c) Concept of an alternate realization, based on Eq.~\eqref{eq:IO_Phen_rot}.  The ferrite Faraday rotator is now removed and reciprocity is now broken by the resonant cylinder mechanically rotating at rate $\Omega$ relative to the waveguide junction.}\label{fig:TCirc}
\end{figure*}

Our circulator approach is conceptually related to four-port ferrite turnstile circulators \cite{Alle56, Auld59}.  Originally developed in the 1950s, these microwave circulators consist of a junction of four rectangular waveguides coupled to a cylindrical resonator~\cite{Pozar}, Fig.~\ref{fig:TCirc}a.  As is typical of most practical circulators, turnstile circulators are rotationally symmetric and preserve the carrier frequencies of signals \cite{Auld59}.  Nonreciprocity comes from a ferrite Faraday rotator in the resonator that rotates the polarization of signals in the cylinder.  For a properly chosen cylinder length and rotation rate, the junction acts as a four port circulator with a finite bandwidth.  That is, when matched loads are placed on the four wave guide ports, microwave signals within the circulator bandwidth are scattered by the junction according to the scattering matrix
\begin{equation}\label{eq:Scirc}
\left[\begin{array}{c}b_{1,\text{out}}(t)\\b_{2,\text{out}}(t)\\b_{3,\text{out}}(t)\\b_{4,\text{out}}(t)\end{array}\right] = \left[\begin{array}{cccc}0&0&0&1\\1&0&0&0\\0&1&0&0\\0&0&1&0\end{array}\right]\left[\begin{array}{c}b_{1,\text{in}}(t)\\b_{2,\text{in}}(t)\\b_{3,\text{in}}(t)\\b_{4,\text{in}}(t)\end{array}\right] 
\end{equation}
where $b_{i,\text{in}}(t)$ and $b_{i,\text{out}}(t)$ are, respectively, the complex envelopes of a traveling wave incident on and scattered by the junction at port $i$.

The circulator's operation is quite modular.  If there were no coupling between the resonator and the waveguides, the scattering between the waveguide ports would be reciprocal
\begin{equation}\label{eq:Stee}
\left[\begin{array}{c}b_{1,\text{out}}(t)\\b_{2,\text{out}}(t)\\b_{3,\text{out}}(t)\\b_{4,\text{out}}(t)\end{array}\right] = \frac12\left[\begin{array}{cccc}-1&1&1&1\\1&-1&1&1\\1&1&-1&1\\1&1&1&-1\end{array}\right]\left[\begin{array}{c}b_{1,\text{in}}(t)\\b_{2,\text{in}}(t)\\b_{3,\text{in}}(t)\\b_{4,\text{in}}(t)\end{array}\right] 
\end{equation}
with a relatively large bandwidth.  This scattering matrix simply corresponds to the signal transfer between an input source driving three loads in parallel, each with the same impedance as the source~\cite{Pozar}.  As described in more detail in section~\ref{sec:Circuit}, when the resonator is coupled to the waveguides, the equation of motion for the input, output, and the two resonant polarization mode envelopes (assuming a signal frequency at the resonator center frequency) is approximately \cite{Yurk84,QN,Goug10,Cler10}
\begin{widetext}\begin{equation}\label{eq:IO_Phen}
{\arraycolsep=1.4pt\def\arraystretch{1.5}\left[\begin{array}{c} \frac{d}{dt}a_x(t)\\ \frac{d}{dt}a_y(t)\\\hline b_{1,\text{out}}(t)\\ b_{2,\text{out}}(t)\\ b_{3,\text{out}}(t)\\ b_{4,\text{out}}(t)\end{array}\right] =  \left[\begin{array}{cc|cccc}-\frac\kappa2&-\Omega&\sqrt{\frac{\kappa}{2}}&0&-\sqrt{\frac{\kappa}{2}}&0\\
\Omega&-\frac\kappa2&0&\sqrt{\frac{\kappa}{2}}&0&-\sqrt{\frac{\kappa}{2}}\\\hline
\sqrt{\frac{\kappa}{2}} & 0 & -\frac12&\frac12&\frac12&\frac12\\
0 &\sqrt{\frac{\kappa}{2}}&\frac12&-\frac12&\frac12&\frac12\\
-\sqrt{\frac{\kappa}{2}}&0&\frac12&\frac12&-\frac12&\frac12\\
0&-\sqrt{\frac{\kappa}{2}}&\frac12&\frac12&\frac12&-\frac12\end{array}\right]\left[\begin{array}{c} a_x(t)\\ a_y(t)\\\hline b_{1,\text{in}}(t)\\ b_{2,\text{in}}(t)\\ b_{3,\text{in}}(t)\\ b_{4,\text{in}}(t)\end{array}\right]}
\end{equation}\end{widetext}    
where $a_{x,y}$ are the complex envelopes of the resonant polarization modes (with units of (photon number)$^{1/2}$), $\kappa$ is the total energy decay rate of a resonator mode into the terminated waveguides, and $\Omega$ is the rate of polarization mixing due to the Faraday rotator.  Thus, the output signals, $b_{i,\text{out}}(t)$ (with units of (photon number/sec)$^{1/2}$), are a linear combination of ``prompt'' scattering from the input signals (as determined by the lower right matrix block in Eq.~\eqref{eq:IO_Phen}) and the resonant modes (lower left matrix block).  The resonant modes are governed by inhomogeneous, first order ordinary differential equations, determined by the modes' ``internal dynamics'' (upper left matrix block), and driving from the input signals (upper right matrix block).  Note that in the absence of the rotator, $\Omega=0$, the $x$- and $y$-polarized resonator modes couple only to waveguides 1 \& 3, and 2 \& 4, respectively.  Thus, orthogonal waveguides (e.g. waveguides 1 and 2) couple only through the prompt scattering.  With the rotator present, excitations are ``rotated'' between the polarization modes, coupling all four waveguides through the resonator as well as through prompt scattering.  

Eq.~\eqref{eq:IO_Phen} can be solved for the steady state response of the resonator modes to input signals.  Doing so, one finds that the polarization of the steady state excitation in the resonator is rotated by an angle $\psi = \text{atan}(2\Omega/\kappa)$, Fig.~\ref{fig:TCirc}b.  When $\psi = \pi/4$, this steady state excitation couples back to each waveguide with equal magnitude.  However, the prompt scattering and resonator-mediated paths carrying signals from $\{1\rightarrow2,2\rightarrow3,3\rightarrow4,4\rightarrow1\}$ interfere constructively, while all other signal path interferences are completely destructive.  Thus, one finds that driving the junction on the cavity's resonance, in steady state, and for $\psi=\pi/4$, the scattering between inputs and outputs becomes exactly Eq.~\eqref{eq:Scirc}.

\begin{figure*}
\includegraphics[width=1\textwidth]{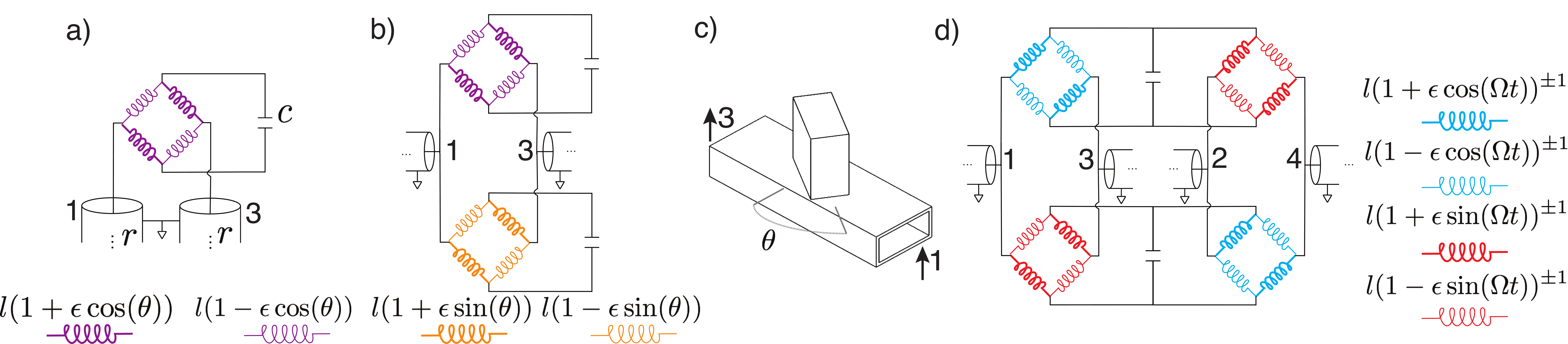}\\
\caption{a) Electrical schematic of a Wheatstone bridge-based LC resonator that serves as the basic module of our four-port circulator design.  Analogous to the turnstile-type structure depicted in c, the resonant mode couples to the two transmission lines 1 \& 3 (characteristic impedance $r$) with a magnitude and sign that depends on $\theta$.  The resonant mode center frequency is $\theta-$independent, which is also true of c.  b) Two such modules placed in parallel, with ``orthogonal'' bridge unbalancing.  d) Four such modules, configured to emulate the mechanically-spinning turnstile depicted in Fig.~\ref{fig:TCirc}c.  As mentioned in the text, the inductances in figure d are modulated linearly in section \ref{sec:Phenomenon}, but the inverse inductances are modulated linearly in section \ref{sec:Circuit}.}\label{fig:TCoupling}
\end{figure*}

However, manipulations of Eq.~\eqref{eq:IO_Phen} suggest alternative ways of realizing the same input-output dynamics.  For example, we can rewrite these dynamics in terms of a rotating basis of the resonator modes
\begin{equation}
\left[\begin{array}{c}a_q(t)\\a_p(t)\end{array}\right] = \left[\begin{array}{cc}\cos(\Omega t)&\sin(\Omega t)\\-\sin(\Omega t)&\cos(\Omega t)\end{array}\right]\left[\begin{array}{c}a_x(t)\\a_y(t)\end{array}\right]
\end{equation}
in which case Eq.~\eqref{eq:IO_Phen} becomes
\begin{widetext}\begin{equation}\label{eq:IO_Phen_rot}
{\arraycolsep=1.4pt\def\arraystretch{1.5}\left[\begin{array}{c} \frac{d}{dt}a_q(t)\\ \frac{d}{dt}a_p(t)\\\hline b_{1,out}(t)\\ b_{2,out}(t)\\ b_{3,out}(t)\\ b_{4,out}(t)\end{array}\right] =  \left[\begin{array}{cc|cccc}-\frac\kappa2&0&\sqrt{\frac{\kappa}{2}}\cos(\Omega t)&\sqrt{\frac{\kappa}{2}}\sin(\Omega t)&-\sqrt{\frac{\kappa}{2}}\cos(\Omega t)&-\sqrt{\frac{\kappa}{2}}\sin(\Omega t)\\
0&-\frac\kappa2&-\sqrt{\frac{\kappa}{2}}\sin(\Omega t)&\sqrt{\frac{\kappa}{2}}\cos(\Omega t)&\sqrt{\frac{\kappa}{2}}\sin(\Omega t)&-\sqrt{\frac{\kappa}{2}}\cos(\Omega t)\\\hline
\sqrt{\frac{\kappa}{2}}\cos(\Omega t) & -\sqrt{\frac{\kappa}{2}}\sin(\Omega t) & -\frac12&\frac12&\frac12&\frac12\\
\sqrt{\frac{\kappa}{2}}\sin(\Omega t) &\sqrt{\frac{\kappa}{2}}\cos(\Omega t)&\frac12&-\frac12&\frac12&\frac12\\
-\sqrt{\frac{\kappa}{2}}\cos(\Omega t)&\sqrt{\frac{\kappa}{2}}\sin(\Omega t)&\frac12&\frac12&-\frac12&\frac12\\
-\sqrt{\frac{\kappa}{2}}\sin(\Omega t)&-\sqrt{\frac{\kappa}{2}}\cos(\Omega t)&\frac12&\frac12&\frac12&-\frac12\end{array}\right]\left[\begin{array}{c} a_q(t)\\ a_p(t)\\\hline b_{1,in}(t)\\ b_{2,in}(t)\\ b_{3,in}(t)\\ b_{4,in}(t)\end{array}\right].}
\end{equation}\end{widetext}    
Inspection of Eq.~\eqref{eq:IO_Phen_rot} suggests an alternate (albeit impractical) realization of a turnstile circulator.  The Faraday rotator dynamics are absent (i.e. no explicit polarization mixing in the upper left hand matrix block), but the coupling between the waveguides and particular resonator polarization modes ``rotates'' in time.  Thus, one might imagine the same turnstile junction, but with the ferrite rod removed and the cylinder physically spinning at a rate $\Omega$, Fig.~\ref{fig:TCirc}c.  

The general structure of Eq.~\eqref{eq:IO_Phen_rot} deserves comment.  Indeed, it has long been recognized that parametrically-modulating components in an electrical network can make a reciprocal network nonreciprocal \cite{Ande65,Ande66,Kama11,Fang12,Ranz14,Este14}.  These modulated components create frequency-sidebands on signal carriers that serve as extra degrees of freedom to encode a physical location onto the signal (e.g. a signal's ``port of entry'').  Nonreciprocal networks then also require either resonant modes or some other means of creating delay between the modulated components.  However, networks that don't demodulate (i.e. coherently ``erase'') these sidebands when the signal exits the network suffer either from more complicated downstream signal processing, or effective loss if the information in the sidebands is ignored.  The network represented in Eq.~\eqref{eq:IO_Phen_rot} achieves perfect demodulation by limiting the signal modulation to the coupling between resonant modes (with no internal loss) and the itinerant fields.  Thus, when signals pass through the resonant modes they are modulated exactly twice, as they enter and as they exit the resonant modes.  In general, only 0$^\text{th}$ and 2$^\text{nd}$ order harmonics of the modulation frequency can appear in the output signal, and it is easy to design these modulations such the 2$^\text{nd}$ order harmonics completely cancel out.  In contrast, many other, related proposals modulate the coupling between resonant modes or between itinerant fields, or some combination of all three modulation types \cite{Ande65,Ande66,Kama11,Fang12,Ranz14,Este14}.  Unfortunately, these types of modulation create sidebands at all harmonic orders, which are more difficult to cancel or to filter out.  

Our approach emulates the dynamics in Eq.~\eqref{eq:IO_Phen_rot}, but as a lumped element, superconducting microwave network with simulated ``spinning'' of two resonant modes relative to four transmission line ports.  Our network can be built up piecewise, starting with the network depicted in Fig.~\ref{fig:TCoupling}a.  In the limit of small inductors (relative to the port impedance) the two transmission lines 1 and 3 are effectively shorted together through the bridge of inductors over most frequencies.  However, the bridge also presents a total inductance of $l$ to the capacitor, forming a resonator with center frequency $1/\sqrt{lc}$.   For $\cos(\theta)\neq0$, the bridge unbalances, coupling the resonant mode to the transmission lines with a magnitude and sign that depends on $\theta$~\cite{Berg10,Abdo14}.  The analogous turnstile-type structure is depicted in Fig.~\ref{fig:TCoupling}c where a two-port rectangular waveguide couples to a rectangular box resonator \cite{Pozar}.  Assuming that only one polarization mode of the box is near resonant with signals applied at the ports, the magnitude and phase of the coupling depend on the orientation angle $\theta$ of the box relative to the waveguide.  Adding a second bridge resonator in parallel with the first, but now with the inductors imbalanced as $\pm\sin(\theta)$ , Fig.~\ref{fig:TCoupling}b, would be analogous to replacing the rectangular box resonator in Fig.~\ref{fig:TCoupling}c with a rotationally symmetric, cylindrical resonator (like that in Fig.~\ref{fig:TCirc}c).    Finally, the full circulator depicted in Fig.~\ref{fig:TCirc}c is emulated by adding two more transmission line ports 2 \& 4 that couple to the same two resonant modes, but do so ``orthogonally'' relative to ports 1 \& 3 via two more bridges.  If the unbalancing of the bridges can be varied sinusoidally in time, i.e. make $\theta = \Omega t$ as in Fig.~\ref{fig:TCoupling}d, then the two resonant modes couple and uncouple from the four transmission lines in a coordinated fashion that simulates the mechanical spinning of the cylinder in the four port circulator depicted in Fig.~\ref{fig:TCirc}c and in Eq.~\eqref{eq:IO_Phen_rot}.  The trick, of course, is to find an experimentally convenient way to both realize dynamically tunable inductors in a superconducting microwave circuit and tune them naturally in such a highly coordinated fashion.    

\begin{figure}[h]
\includegraphics[width=.5\textwidth]{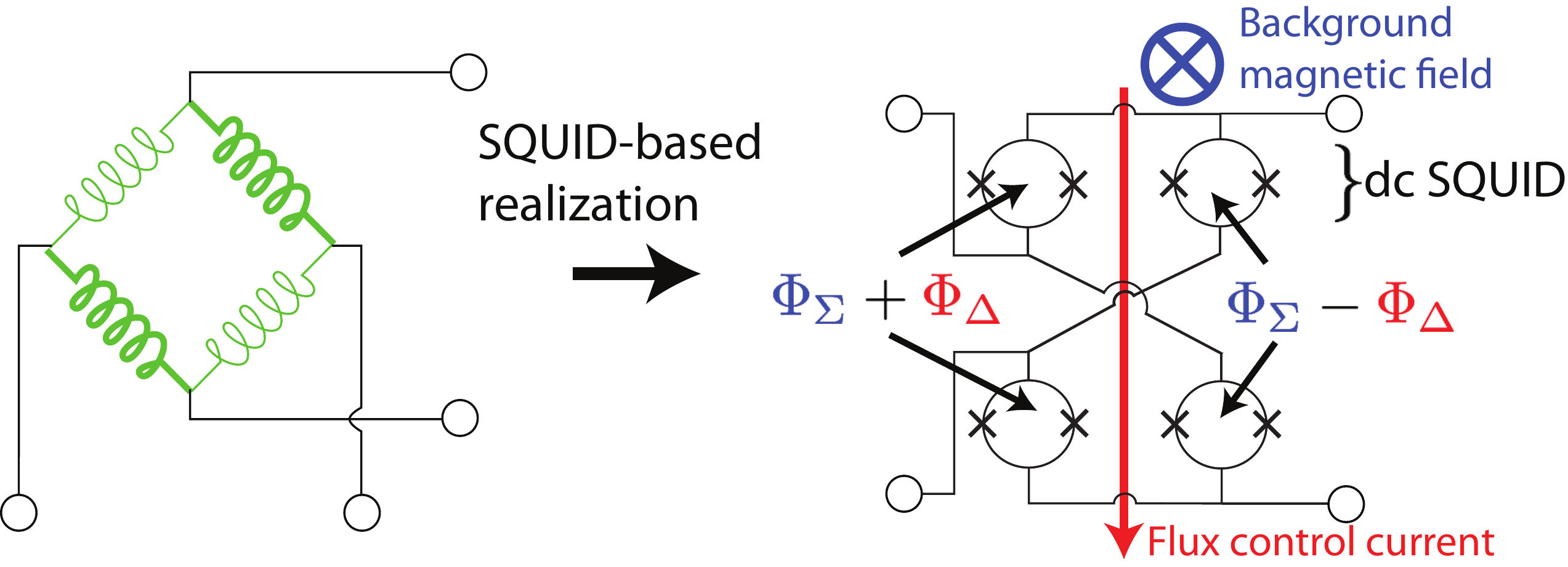}\\
\caption{Schematic of a dc SQUID-based realization of the tunable bridge network.  In the right hand figure, the wire configuration (black lines) is critical as it determines the magnetic flux through the five depicted loops (four SQUID loops, one loop of the bridge itself.  The four small circles represent unimportant network terminals.).  The total magnetic flux $\Phi$ through each SQUID loop is the sum of the flux from a uniform background field ($\Phi_\Sigma$) and from a magnetic flux control current that flows vertically through the ``twisted'' bridge ($\Phi_\Delta$).  Ideally, the net magnetic flux through the bridge loop is always zero.}\label{fig:BSchem}
\end{figure}

Our tunable inductors are dc superconducting quantum interference devices (SQUIDs), which are two Josephson junctions connected in parallel~\cite{Duzer}.  For currents much smaller than the SQUID critical current, the SQUID acts as an inductor with inductance
\begin{equation}\label{eq:Ls}
l_s = \frac{\varphi_0}{I_s},\quad I_s=2I_0\left|\cos\left(\frac{\Phi}{2\varphi_0}\right)\right|
\end{equation}   	
(assuming identical junctions and negligible geometric inductance) where $I_s$ is the SQUID critical current, $I_0$ the junction critical current, $\varphi_0$ the reduced flux quantum, and $\Phi$ the magnetic flux threading the SQUID loop.  Replacing the four inductors in Fig.~\ref{fig:TCoupling}a with four SQUIDs realizes a bridge of dynamically tunable inductors.  Furthermore, twisting the bridge into a figure-eight layout gives us a convenient way to achieve the required unbalancing of the bridge inductances.  We can set the inductance of each SQUID appropriately by applying a constant and uniform magnetic field and a smaller, time-dependent, and gradiometic magnetic field, as in Fig.~\ref{fig:BSchem}.  Note, too, that this configuration only induces the desired screening currents within each SQUID loop.  Undesired screening currents between SQUIDs are not induced (canceling unwanted screening currents in a network like Fig.~\ref{fig:TCoupling}d is also important, but outside the scope of this article).  Thus, in principle, all of the inductive, dynamically unbalanced bridges in Fig.~\ref{fig:TCoupling}d, could be realized with 16 SQUIDs, a uniform background magnetic field and two control wires carrying current oscillating as $\cos(\Omega t)$ and $\sin(\Omega t)$.  In section~\ref{sec:SQUID}, we will also address considerations such as increasing the network's saturation power by replacing single SQUIDs with arrays and the secant- rather than linear-dependence of inductance on magnetic flux.    

\section{Circuit analysis}\label{sec:Circuit}

\begin{figure*}
\includegraphics[width=1\textwidth]{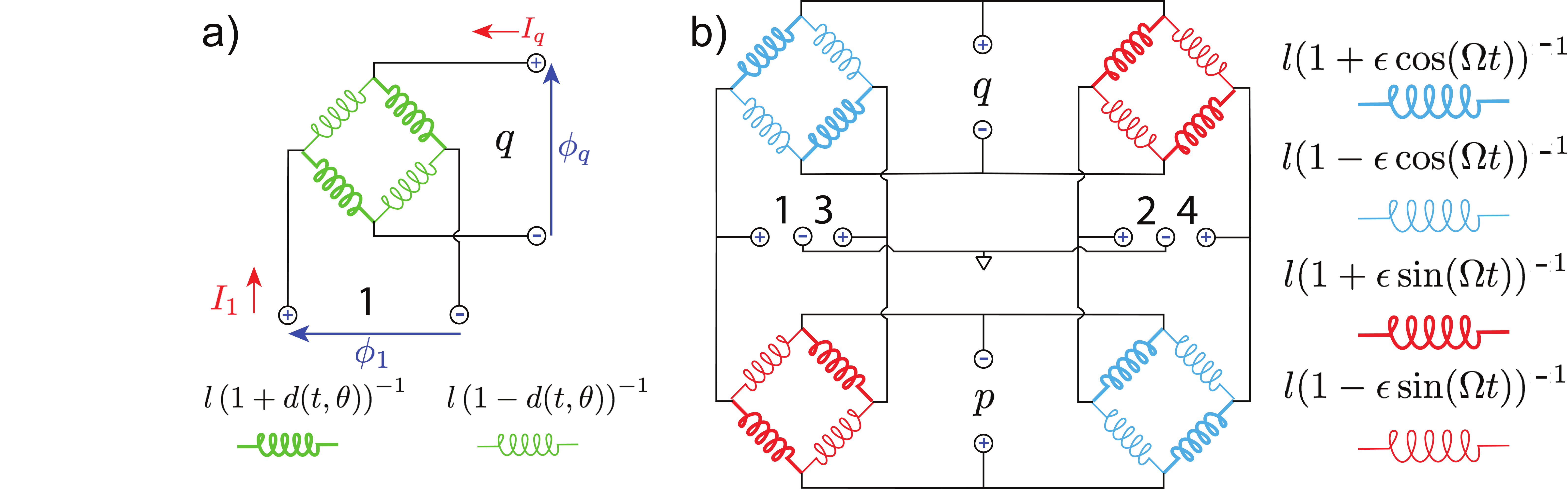}\\
\caption{a) The inductance bridge as a network with two ports, 1 \& $q$.  b)  Four inductance bridges configured as a six port network.  Attaching a capacitor across ports $q$ \& $p$ and transmission lines across ports 1-4 realizes the circulator network depicted in Fig.~\ref{fig:TCoupling}d.}\label{fig:BAnaly}
\end{figure*}

We now analyze the circuit depicted in Fig.~\ref{fig:TCoupling}d.  Analysis will use both the frequency domain, lumped element approach common in microwave engineering~\cite{Pozar}, and a time-domain approach, which shows how Eqs.~\eqref{eq:IO_Phen_rot} approximate the circuit's dynamics.  

The essential sub-network in this circulator is the two port \cite{Pozar} Wheatstone bridge-type \cite{Berg10} network depicted in Fig.~\ref{fig:BAnaly}a.  In contrast to the networks depicted in Fig.~\ref{fig:TCoupling}, in this section we will exclusively consider inductances that vary as $l(1\pm d(t,\theta))^{-1}$ ($l$ some inductance, and $d$ is a real function of time $t$, and an angle $\theta$), as in Fig.~\ref{fig:BAnaly}a.  This is a more natural, simplified model for the dependence of SQUID inductances on magnetic flux, Eq.~\eqref{eq:Ls}.  Straightforward circuit analysis \cite{Pozar} gives us the constitutive equations for this network
\begin{equation}\label{eq:BridgeAdmit}
\frac1l\left[\begin{array}{cc}1&d(t,\theta)\\ d(t,\theta)&1\end{array}\right]\left[\begin{array}{c}\phi_1(t)\\\phi_q(t)\end{array}\right]=\left[\begin{array}{c}I_1(t)\\I_q(t)\end{array}\right].
\end{equation}
where $\phi_i(t) = \int_{-\infty}^tV_{i}(\tau)d\tau$ is the time integral of the voltage across port $i$ and is called the ``branch flux'' across this port~\cite{Devo95}, and $I_i$ is the current entering at port $i$.  Throughout this article, we will use the convention that inductance is defined as the ratio of the branch flux across and current through a network branch \cite{Duzer,Yurk89} (which for time-varying or nonlinear inductances is different from the more familiar definition of inductance as the ratio of voltage and the time derivative of current).  From Eq.~\eqref{eq:BridgeAdmit}, we learn that the unbalancing of the bridge in Fig.~\ref{fig:BAnaly}a gives an output reluctance (the current response at one port per flux applied at the other; i.e. inverse inductance) proportional to $d(t,\theta)$, but the input reluctance (the current response at one port per flux applied at the same port) is independent of $d(t,\theta)$.  Such clean separation between input and output reluctances is highly attractive when $d(t,\theta)$ is proportional to an experimentally-convenient tunable parameter, such as the control current in Fig.~\ref{fig:BSchem}.  Our network is designed to exploit this separation to realize time-variable coupling between resonant modes and itinerant fields {\it without} modulating the intra-resonator dynamics or prompt field scattering.  This bridge is the repeated module that helps us realize Eq.~\eqref{eq:IO_Phen_rot}.    

Next, Fig.~\ref{fig:BAnaly}b takes four copies of this bridge, places them in a ring, and defines six ports $\{1,2,3,4,q,p\}$.  Note that ports 1-4 are each defined between a network node and ground, while ports $q$ and $p$ are each defined by two network nodes.  If we let $d(t,\theta) = \epsilon\cos(\Omega t+\theta)$ ($0\leq\epsilon\leq1$ ), straightforward circuit analysis then relates port branch fluxes and currents:
\begin{widetext}\begin{equation}\label{eq:LNet}
\frac1l\left[
\begin{array}{cc|cccc}
 2 & 0 & \epsilon  \cos (\Omega t ) & \epsilon  \sin (\Omega t ) &
   -\epsilon  \cos (\Omega t ) & -\epsilon  \sin (\Omega t ) \\
 0 & 2 & \epsilon  \sin (\Omega t ) & -\epsilon  \cos (\Omega t ) &
   -\epsilon  \sin (\Omega t ) & \epsilon  \cos (\Omega t ) \\\hline
 \epsilon  \cos (\Omega t ) & \epsilon  \sin (\Omega t ) & 3 & -1 & -1 &
   -1 \\
 \epsilon  \sin (\Omega t ) & -\epsilon  \cos (\Omega t ) & -1 & 3 & -1 &
   -1 \\
 -\epsilon  \cos (\Omega t ) & -\epsilon  \sin (\Omega t ) & -1 & -1 & 3 &
   -1 \\
 -\epsilon  \sin (\Omega t ) & \epsilon  \cos (\Omega t ) & -1 & -1 & -1 &
   3 \\
\end{array}
\right]\left[\begin{array}{c}\phi_q(t)\\\phi_p(t)\\\hline\phi_1(t)\\\phi_2(t)\\\phi_3(t)\\\phi_4(t)\end{array}\right]=\left[\begin{array}{c}I_q(t)\\I_p(t)\\\hline I_1(t)\\I_2(t)\\I_3(t)\\I_4(t)\end{array}\right].
\end{equation}\end{widetext}
While the structure of Eq.~\eqref{eq:LNet} is evocative of Eq.~\eqref{eq:IO_Phen_rot}, they are not equivalent.  For example, Eq.~\eqref{eq:LNet} contains no resonant dynamics yet and its dynamics are reciprocal \cite{Ande65,Pozar}.  Nonetheless, we already have the critical structure that only the couplings between ports 1-4 and ports $q$ \& $p$ are variable in time.  Analysis of Eq.~\eqref{eq:LNet} simplifies by going into a ``left-right/even-odd'' basis for the port 1-4 variables and a rotating, circular basis for the $q$ and $p$ port variables:   
\begin{eqnarray}\label{eq:EO_basis}
\left[\begin{array}{c}\phi_{l,e}(t)\\\phi_{r,e}(t)\\\phi_{l,o}(t)\\\phi_{r,o}(t)\end{array}\right] &=&\frac{1}{\sqrt{2}}\left[\begin{array}{cccc}1&0&1&0\\
0&1&0&1\\
1&0&-1&0\\
0&1&0&-1\end{array}\right] \left[\begin{array}{c}\phi_{1}(t)\\\phi_{2}(t)\\\phi_{3}(t)\\\phi_{4}(t)\end{array}\right],\\
\left[\begin{array}{c}\phi_{+}(t)\\\phi_{-}(t)\end{array}\right] &=&\frac{1}{\sqrt{2}}\left[\begin{array}{cccc}e^{j\Omega t}&-je^{j\Omega t}\\
e^{-j\Omega t}&je^{-j\Omega t}\end{array}\right] \left[\begin{array}{c}\phi_{q}(t)\\\phi_{p}(t)\end{array}\right]
\end{eqnarray}
and similarly for the current variables ($j$ being the imaginary unit, adopting the electrical engineering convention~\footnote{In which the Fourier transform of a function is defined as $F[\omega] = \frac{1}{\sqrt{2\pi}}\int_{-\infty}^{\infty}F(t)e^{-j t}dt$}).  In these bases, the dynamics of Eq.~\eqref{eq:LNet} separate into:
\begin{eqnarray}\label{eq:LNet_E}
\frac2l\left[
\begin{array}{cc}
1&-1\\
-1&1
\end{array}
\right]\left[\begin{array}{c}\phi_{l,e}(t)\\\phi_{r,e}(t)\end{array}\right]&=&\left[\begin{array}{c}I_{l,e}(t)\\I_{r,e}(t)\end{array}\right],\\ \label{eq:LNet_O}
\frac1l\left[
\begin{array}{cc|cc}
 2 & 0 & \epsilon& j\epsilon  \\
 0 & 2 & \epsilon & -j\epsilon  \\\hline
\epsilon & \epsilon & 4 & 0  \\
-j\epsilon & j\epsilon & 0 & 4 
\end{array}
\right]\left[\begin{array}{c}\phi_+(t)\\\phi_-(t)\\\hline\phi_{l,o}(t)\\\phi_{r,o}(t)\end{array}\right]&=&\left[\begin{array}{c}I_+(t)\\I_-(t)\\\hline I_{l,o}(t)\\I_{r,o}(t)\end{array}\right].
\end{eqnarray}

The time-dependent, reciprocal network of Fig.~\ref{fig:BAnaly}b becomes nonreciprocal when we make modes $q$ and $p$ resonant by placing a capacitor (capacitance $c$) across each of these ports.  Doing so, these ports gain a fixed relation between current and branch flux
\begin{eqnarray}
-c\frac{d^2}{dt^2}\phi_{q,p}(t) &=& I_{q,p}(t),
\end{eqnarray}
or equivalently in the circular basis
\begin{eqnarray}
-c\left(\frac{d}{dt}\mp j\Omega\right)^2\phi_{\pm}(t)&=&I_\pm(t).
\end{eqnarray}
The capacitors have no effect on the ``even'' dynamics in Eq.~\eqref{eq:LNet_E}, but in the following subsection we will show that they turn the reciprocal, four-port ``odd'' network Eq.~\eqref{eq:LNet_O} into a nonreciprocal, two port network.

\subsection{Frequency-domain analysis}\label{sec:freq}

Taking into account the capacitors and writing the odd network dynamics now in the frequency domain, we find:
\begin{eqnarray}\label{eq:Gyr_Freq}
\frac1l\left[
\begin{array}{cc|cc}
 2-lc(\omega-\Omega)^2 & 0 & \epsilon& j\epsilon  \\
 0 & 2-lc(\omega+\Omega)^2 & \epsilon & -j\epsilon  \\\hline
\epsilon & \epsilon & 4 & 0  \\
-j\epsilon & j\epsilon & 0 & 4 
\end{array}
\right]\left[\begin{array}{c}\phi_{+}[\omega]\\\phi_{-}[\omega]\\\hline\phi_{l,o}[\omega]\\\phi_{r,o}[\omega]\end{array}\right]=&&\nonumber\\
\left[\begin{array}{c}0\\0\\\hline I_{l,o}[\omega]\\I_{r,o}[\omega]\end{array}\right]&&
\end{eqnarray}
where square brackets $[\cdot]$ indicate a frequency domain variable.  From Eq.~\eqref{eq:Gyr_Freq}, one finds that a current applied at either the ``left'' or ``right'' odd ports (i.e. taking $I_{l,o}\neq0$ or $I_{r,o}\neq0$) induces a resonant response in the $\phi_\pm$ variables when $\omega =  \sqrt{(4-\epsilon^2)/2lc}\pm\Omega\equiv\omega_0\pm\Omega$, respectively.  For $\Omega=0$, the orthogonal coupling between the odd modes and the resonant modes means that the left and right odd modes are uncoupled over all drive frequencies.  But, for $\Omega\neq0$ and odd driving at the frequency $\omega_0$, the $\phi_+$ and $\phi_-$ modes respond as equal and opposite capacitive and inductive reactances in parallel, which open a transmission window between the left and right odd ports \cite{Fay65}.  For instance, eliminating $\phi_\pm$ in Eq.~\eqref{eq:Gyr_Freq}, setting $\omega=\omega_0$, and expanding to first order in $\Omega$ we find that 
\begin{eqnarray}\label{eq:Gyr_Freq_O0}
j\omega_0\frac{16c\Omega}{\epsilon^2}\left[
\begin{array}{cc}
0&1\\-1&0
\end{array}
\right]\left[\begin{array}{c}\phi_{l,o}[\omega_0]\\\phi_{r,o}[\omega_0]\end{array}\right]=
\left[\begin{array}{c}I_{l,o}[\omega_0]\\I_{r,o}[\omega_0]\end{array}\right].
\end{eqnarray}  
In this configuration, signal transfer through the network is nonreciprocal: voltage applied at the right odd port will induce a positive current at the left odd port, but voltage applied at the left odd port will induce a negative current at the left odd port (note that $j\omega_0$ times a branch flux is a voltage).  A network with the constituent relations given in Eq.~\eqref{eq:Gyr_Freq_O0} is known as a ``gyrator,'' the most basic building block of nonreciprocity in electrical network theory, with ``gyration resistance'' $\epsilon^2/16c\Omega$ \cite{Tell48}.  

Further analysis is aided by attaching transmission lines to ports 1-4.  With the capacitors across ports $p$ and $q$ and transmission lines across ports 1-4, Fig.~\ref{fig:BAnaly}b becomes equivalent to Fig.~\ref{fig:TCoupling}d.  The voltages and currents at ports 1-4 are now decomposed into traveling wave voltages and currents incident upon (``in'') and scattered by (``out'') the network through the transmission lines \cite{Pozar, Yurk84, Goug10}.  Our goal is now to relate these input and output waves.  While such input-output analysis is convenient theoretically when studying four-port circulator networks~\footnote{This is because an ideal four port network doesn't always have an admittance or impedance matrix representation \cite{Fay65}.}, measuring scattering parameters is also more experimentally convenient in microwave networks than direct measurements of port voltages and currents.   

If the only constraints on the input and output fields are those imposed by our network (e.g. the transmission lines are of effectively infinite length) then \cite{Pozar,Yurk84,Cler10}
\begin{eqnarray}\label{eq:IO_time}
I_i(t) &=& I_{i,\text{in}}(t)+I_{i,\text{out}}(t)\nonumber\\
\frac{d}{dt}\phi_i(t)&=& V_{i,\text{in}}(t)+V_{i,\text{out}}(t)\nonumber\\
r I_{i,\text{in}}(t)&=& V_{i,\text{in}}(t)\nonumber\\
-r I_{i,\text{out}}(t)&=& V_{n,\text{out}}(t)
\end{eqnarray}
where the traveling wave variables above are evaluated at the port positions and $r$ is the characteristic (real) impedance of each transmission line.  Standard network analysis \cite{Pozar} then tells us that if the port branch fluxes and currents are related by 
\begin{equation}
j\omega\mathbb{Y}_{ij}[\omega]\phi_j[\omega] = I_i[\omega] 
\end{equation}
where $\mathbb{Y}$ is known as the admittance matrix, it then follows that
\begin{eqnarray}\label{eq:S_admit}
V_{i,\text{out}}[\omega] &=& \left[(\mathbb{1}+r\mathbb{Y}[\omega])^{-1}(\mathbb{1}-r\mathbb{Y}[\omega])\right]_{ij} V_{j,\text{in}}[\omega]\nonumber\\
&\equiv&\mathbb{S}_{ij}[\omega]V_{j,\text{in}}[\omega], 
\end{eqnarray}
where $\mathbb{1}$ is the identity and provided that the matrix inverse exists.  $\mathbb{S}$ is often referred to as a scattering matrix.  

For example, using Eq.~\eqref{eq:Gyr_Freq_O0}, one finds that when $\Omega=\Omega_0\equiv\epsilon^2/16cr$
\begin{equation}\label{eq:Gyr_Scatter_admit}
\left[\begin{array}{c}V_{l,o,\text{out}}[\omega_0]\\V_{r,o,\text{out}}[\omega_0]\end{array}\right]= \left[\begin{array}{cc}0&-1\\1&0\end{array}\right]\left[\begin{array}{c}V_{l,o,\text{in}}[\omega_0]\\V_{r,o,\text{in}}[\omega_0]\end{array}\right].
\end{equation}
In other words, when the gyrator resistance equals $r$, the odd network is matched: signals are fully transmitted by the network, but the transmission is nonreciprocal.  A voltage signal picks up a $\pi$ phase shift traveling from the right odd to left odd port and no phase shift traveling from left odd to right odd.  Similarly, using the even network relations in Eq.~\eqref{eq:LNet_E}, one finds that to zeroth order in $\omega_0 l/r$
\begin{equation}\label{eq:Even_Scatter_admit}
\left[\begin{array}{c}V_{l,e,\text{out}}[\omega_0]\\V_{r,e,\text{out}}[\omega_0]\end{array}\right]= \left[\begin{array}{cc}0&1\\1&0\end{array}\right]\left[\begin{array}{c}V_{l,e,\text{in}}[\omega_0]\\V_{r,e,\text{in}}[\omega_0]\end{array}\right].
\end{equation}
Combining Eqs.~\eqref{eq:Gyr_Scatter_admit} and~\eqref{eq:Even_Scatter_admit} and putting the input and output voltage signals back into the port 1-4 basis, we find that this network acts as an ideal, four-port circulator, Eq.~\eqref{eq:Scirc}, for input voltage waves at frequency $\omega_0$.  We also mention that as our network is operable as a two port gyrator, it is therefore also operable as a three-port circulator by combining ports 3 and 4 in Fig.~\ref{fig:TCoupling}d into a single port (i.e. by merging the transmission lines 3 and 4 together) \cite{Viol14,Carl55}.

\subsection{Time-domain analysis}\label{sec:time}

This standard scattering matrix approach, though, obscures the dynamics that guided our intuition: a simulated ``rotation'' of the resonant modes relative to the itinerant fields (e.g. Fig.~\ref{fig:TCoupling}c).  To that end, we reanalyze the dynamics of Eq.~\eqref{eq:LNet_E} and Eq.~\eqref{eq:Gyr_Freq} in the time-domain, and without eliminating the $\phi_\pm$ dynamics.  In quantum optics, analogous models are known as ``input-output'' (IO) models~\cite{QN,Goug10,Yurk84,Cler10}.  Like the scattering matrix models of the previous paragraph, IO models relate incident and scattered electromagnetic waves and are thus natural models for resonant microwave networks.  But unlike scattering matrix models, they are time-domain, first order ordinary differential equations.  IO models approximate resonant circuit dynamics, but may also be directly compared with, for example, optical systems that don't have lumped element representations.  On the other hand, the circuit analysis in the previous section does not {\it require} approximations beyond the lump element assumptions, and can be useful in considering how IO approximations break down.     

In this case, our IO model assumes that the network variables have solutions of the form \cite{QN, Yurk84,Cler10}
\begin{eqnarray}\label{eq:sin_sol_IO}
\phi_{i}(t) &=& \varphi_{i}(t)e^{j\omega_d t}+\mathrm{c.c.}\nonumber\\
V_{i,\text{in}}(t) &=& \omega_0\sqrt{cr}v_{i,\text{in}}(t)e^{j\omega_d t}+ \mathrm{c.c.}\nonumber\\
V_{i,\text{out}}(t) &=& \omega_0\sqrt{cr}v_{i,\text{out}}(t)e^{j\omega_d t}+ \mathrm{c.c.}\nonumber\\
\omega_d &=& \omega_0+\Delta.
\end{eqnarray}
for a drive frequency detuned by $\Delta$ from the center frequency $\omega_0$.  Then, combining Eqs.~\eqref{eq:LNet_E},~\eqref{eq:Gyr_Freq}, and~\eqref{eq:IO_time}, making a slowly varying envelope approximation and solving to lowest order in $\omega_0l/r$, $|\Delta|/\omega_0$, and $|\Omega|/\omega_0$ ({\it i.e.} assuming that the rate of variation of $\varphi_i,$ $v_{i,\text{in}}$, and $v_{i,\text{out}}$ is much less than $\omega_d$) gives
\begin{widetext}\begin{eqnarray}\label{eq:Gyr_rot_IO}
\left[\begin{array}{c}\frac{d}{dt}\varphi_+(t)\\\frac{d}{dt}\varphi_-(t)\\\hline v_{l,o,\text{out}}(t)\\v_{r,o,\text{out}}(t)\end{array}\right] = \left[\begin{array}{cc|cc}-\left(j(\Delta-\Omega)+\frac{\kappa}{2}\right)&0&j\sqrt{\frac{\kappa}{2}}&-\sqrt{\frac{\kappa}{2}}\\
0&-\left(j(\Delta+\Omega)+\frac{\kappa}{2}\right)&j\sqrt{\frac{\kappa}{2}}&\sqrt{\frac{\kappa}{2}}\\\hline
-j\sqrt{\frac{\kappa}{2}}&-j\sqrt{\frac{\kappa}{2}}&-1&0\\
-\sqrt{\frac{\kappa}{2}}&\sqrt{\frac{\kappa}{2}}&0&-1\end{array}\right]\left[\begin{array}{c}\varphi_+(t)\\\varphi_-(t)\\\hline v_{l,o,\text{in}}(t)\\v_{r,o,\text{in}}(t)\end{array}\right],\,\left[\begin{array}{c}v_{l,e,\text{out}}(t)\\v_{r,e,\text{out}}(t)\end{array}\right] = \left[\begin{array}{cc}0&1\\1&0\end{array}\right]\left[\begin{array}{c}v_{l,e,\text{in}}(t)\\v_{r,e,\text{in}}(t)\end{array}\right]
\end{eqnarray}\end{widetext}
where we have defined $\kappa = \epsilon^2/8cr$.  We call Eq.~\eqref{eq:Gyr_rot_IO} an IO model of the network depicted in Fig.~\ref{fig:TCoupling}d.  Putting this model back into the $\{1,2,3,4,q,p\}$ basis then gives us the four port circulator model Eq.~\eqref{eq:IO_Phen_rot} with the identifications $\{a_q,a_p,b_{1,\text{(in/out)}},b_{2,\text{(in/out)}},b_{3,\text{(in/out)}},b_{4,\text{(in/out)}}\}=\{\varphi_q,\varphi_p,v_{1,\text{(in/out)}},v_{2,\text{(in/out)}},v_{3,\text{(in/out)}},v_{4,\text{(in/out)}}\}$.  A natural interpretation of the IO model Eq.~\eqref{eq:IO_Phen_rot} is that while the center frequency and total damping of each resonator mode are constant in time, the resonant modes couple to (and are damped by) each transmission line in a sinusoidally ``rotating'' pattern.  While the rotating coupling between resonant and port degrees of freedom is apparent in Fig.~\ref{fig:TCoupling}d, the time-independent intra-resonator dynamics are less obvious. 

Finally, we calculate the performance of the circulator in this IO representation in order to compare it to the frequency domain circuit analysis presented in section~\ref{sec:freq}.  For linear IO models of the form
\begin{eqnarray}
\left[\begin{array}{c}\frac{d}{dt}\vec{x}(t)\\\hline \vec{y}_{\text{out}}(t)\end{array}\right] = \left[\begin{array}{c|c}\mathbb{A}&\mathbb{B}\\\hline\mathbb{C}&\mathbb{D}\end{array}\right]\left[\begin{array}{c}\vec{x}(t)\\\hline \vec{y}_{\text{in}}(t)\end{array}\right]
\end{eqnarray}
with time-independent matrices $\{\mathbb{A},\mathbb{B},\mathbb{C},\mathbb{D}\}$, a scattering matrix-like representation of the steady state response of the output amplitude vector $\vec{y}_{\text{out}}$ to a constant input amplitude vector $\vec{y}_{\text{out}}$ is 
\begin{eqnarray}\label{eq:Gyr_IO_Scat}
\vec{y}_{\text{out}}^{ss} &=& \left(\mathbb{D}-\mathbb{C}\mathbb{A}^{-1}\mathbb{B}\right)\vec{y}_{\text{in}}^{ss}
\end{eqnarray}
in general.  However, as is typical for IO models of lossless, gain-less, linear systems, Eqs.~\eqref{eq:Gyr_rot_IO} have a particular form where $\mathbb{B} = -\mathbb{C}^\dag \mathbb{D}$, $\mathbb{A} = -j\mathbb{Q}-\frac12\mathbb{C}^\dag\mathbb{C}$, and $\mathbb{D}$ is unitary and $\mathbb{Q}$ is Hermitian ($^\dag$ signifying the matrix adjoint) \cite{Goug10}.  In this case, in analogy with Eq.~\eqref{eq:S_admit}, it can be shown that \cite{Gough} 
\begin{eqnarray}\label{eq:S_io}
\vec{y}_{\text{out}}^{ss} &=& \mathbb{D}^{1/2}\left(\mathbb{1}+\mathbb{Y}_{\text{io}}\right)^{-1}\left(\mathbb{1}-\mathbb{Y}_{\text{io}}\right)\mathbb{D}^{1/2}\vec{y}_{\text{in}}^{ss}\nonumber\\
\mathbb{Y}_{\text{io}}&=&2j\mathbb{D}^{1/2}\left(\mathbb{C}\mathbb{Q}^{-1}\mathbb{C}^\dag\right)^{-1}\mathbb{D}^{-1/2}
\end{eqnarray}
(assuming the matrix inverses exist).  Using Eq.~\eqref{eq:S_io} and the first Eq. in~\eqref{eq:Gyr_rot_IO}, we find an admittance-like matrix for the IO model of the odd modes 
\begin{equation}
\mathbb{Y}_{\text{io},\text{odd}} = \frac{-2}{\kappa}\left[\begin{array}{cc}j\Delta & -\Omega\\\Omega&j\Delta\end{array}\right].
\end{equation}
And thus, in analogy with Eq.~\eqref{eq:Gyr_Scatter_admit}, we find for $\Delta = 0$ and $\Omega = \kappa/2$ (i.e. for $\omega_d=\omega_0$ and $\Omega = \Omega_0$) this IO acts like an ideal gyrator 
\begin{equation}\label{eq:S_IO}
\left[\begin{array}{c}v_{l,e,\text{out}}^{ss}\\v_{r,e,\text{out}}^{ss}\end{array}\right] = \left[\begin{array}{cc}0&-1\\1&0\end{array}\right]\left[\begin{array}{c}v_{l,e,\text{in}}^{ss}\\v_{r,e,\text{in}}^{ss}\end{array}\right].
\end{equation}
If tuned optimally ($\Omega = \Omega_0$) and for $\Delta \neq 0$, the FWHM bandwidth of the odd and full circulator network models are $\sqrt{2}\kappa$ and $\sqrt{2(\sqrt{3}-1)}\kappa$, respectively.  Thus circulator bandwidth is strongly constrained by the modulation amplitude of the inductors, $\kappa\propto\epsilon^2$, as will be discussed more in section~\ref{sec:SQUID}.  We also note that $\sqrt{2}\kappa=2\sqrt{2}\Omega_0$, which is $\sqrt{2}$ times the frequency splitting of the internal modes for a matched network.  A simple proportionality between bandwidth and the frequency splitting of internal modes is typical of many non-reciprocal networks \cite{Fay65}.  

\begin{figure*}
\includegraphics[width=1\textwidth]{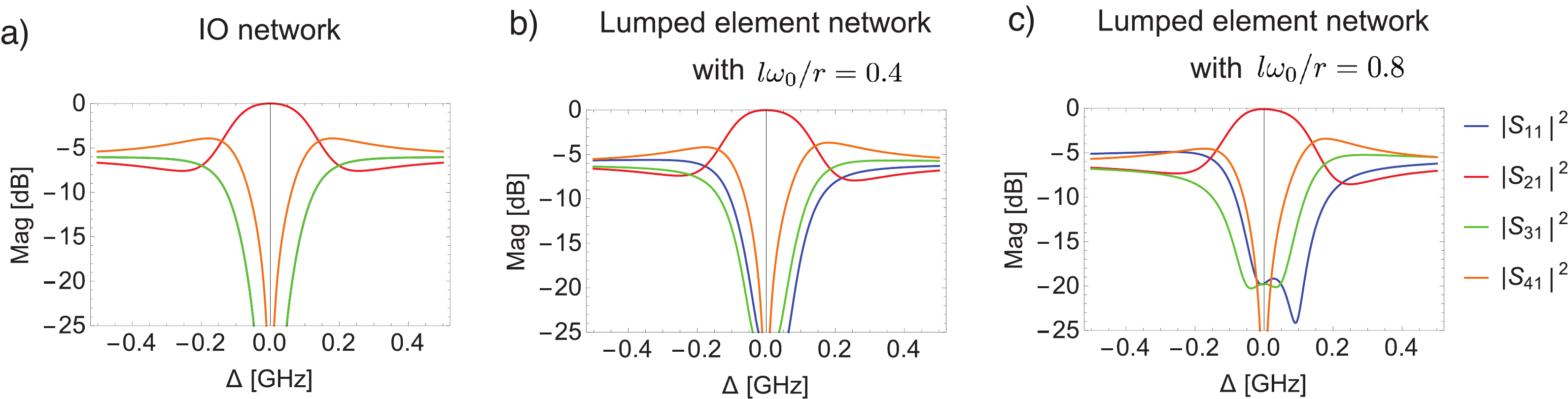}\\
\caption{Simulation of the frequency domain response of various models of the circulator, as described in the text.  Curves of the other scattering parameters are identical to these, with a cyclic permutation of the port indices, confirming proper circulation.}\label{fig:Admit_IO_Circ_Graphs}
\end{figure*}

\subsection{Simulation}

It is worth comparing the theoretical scattering response of the approximate IO model to the full lumped-element scattering matrix (which is exact up to the lumped element assumption).  In the IO model, the scattering response corresponds to an ideal four port circulator when $\Delta=0$ and $\Omega=\kappa/2$.  Alternatively, one can calculate $\mathbb{S}$ for the full lumped element network  depicted in Fig.~\ref{fig:TCoupling}d, but without the approximations made to produce Eqs.~\eqref{eq:Gyr_Scatter_admit}-\eqref{eq:Even_Scatter_admit}.  Choosing parameters compatible with superconducting microwave circuits~\cite{Mall11,Cast08} $l=\frac12$ nH, $c=2$ pF, $r= 50\,\Omega$, $\omega = \omega_0=2\pi\times6.16$ GHz, $\Omega = \Omega_0=2\pi\times99$ MHz, and $\epsilon = 1$ we find:
\begin{eqnarray}\label{eq:S_Power}
|\mathbb{S}_{11}|^2=|\mathbb{S}_{22}|^2=|\mathbb{S}_{33}|^2=|\mathbb{S}_{44}|^2=0.002\nonumber\\
|\mathbb{S}_{21}|^2=|\mathbb{S}_{32}|^2=|\mathbb{S}_{43}|^2=|\mathbb{S}_{14}|^2=0.995\nonumber\\
|\mathbb{S}_{31}|^2=|\mathbb{S}_{42}|^2=|\mathbb{S}_{13}|^2=|\mathbb{S}_{24}|^2=0.002\nonumber\\
|\mathbb{S}_{41}|^2=|\mathbb{S}_{12}|^2=|\mathbb{S}_{23}|^2=|\mathbb{S}_{34}|^2=0.000
\end{eqnarray}  
That is, 99.5\% of the incident power at this frequency is routed correctly for these parameters and 0.2\% is reflected, assuming ideal lumped element components and ideal inductor modulation (section~\ref{sec:SQUID} will focus on the many considerations that go into identifying a practical set of parameters).  The network's response to off-resonant drives, $\Delta=\omega_d-\omega_0\neq0$, is depicted in Fig.~\ref{fig:Admit_IO_Circ_Graphs} for the various models.  For these parameters, the circulator bandwidth is $\sqrt{2(\sqrt{3}-1)}\kappa/2\pi=$ 241 MHz in the IO model response, which is depicted in Fig.~\ref{fig:Admit_IO_Circ_Graphs}a.  We see a very similar response using the lumped element circuit model in Fig.~\ref{fig:Admit_IO_Circ_Graphs}b.  The greatest difference in this response is that that it is no longer perfectly symmetric about $\Delta=0$, especially outside of the circulator bandwidth.  

The discrepancy between the lumped element and IO models appears to be largely due to nonidealities in the even network transmission Eq.~\eqref{eq:Even_Scatter_admit}: perfect circulation breaks down when the even network ceases to look like two ports shorted together.  Thus the ideal even network response occurs in the limit of small $\omega_0l/r$, and for these parameters $\omega_0l/r=0.39$.  If instead we choose the parameters $l=1$ nH, $c=1$ pF, $\epsilon = 1/\sqrt{2}$ ($\omega_0=2\pi\times6.7$ GHz), the corresponding IO model still has the same $\omega_0$, $\kappa$, and $\Omega_0$, but now $\omega_0l/r=0.84$.  Evaluating the lumped element scattering parameters for these values now results in a response that deviates further from the ideal IO model, with a 97.8\% maximum power transmission, 1\% power reflected, and a more asymmetric response about $\Delta=0$, Fig.~\ref{fig:Admit_IO_Circ_Graphs}c.     

\section{Considerations for a SQUID-based realization}\label{sec:SQUID}

To better ground the above analysis in the reality of superconducting microwave circuits, we now outline some considerations that go into designing dc SQUID networks that approximate linear inductors.  Circulator performance should be compared against commercial circulators and the needs of contemporary and future experiments with superconducting quantum microwave circuits.  Desirable characteristics for our circulator include an input-output response that is close to Eq.~\eqref{eq:Scirc} on resonance, a wide bandwidth, and high power handling.  For example, there is a trade-off in this particular circuit design (Fig.~\ref{fig:TCoupling}d) in that ideal input-output response occurs in the limit of small $\omega_0l/r$, subsection~\ref{sec:Circuit}c, but the bandwidth of the network also decreases in this same limit ($\kappa/\omega_0\sim l\omega_0/r$).  This trade-off is independent of any component nonidealities and arises at the schematic level of this particular design.  We also note that competition between bandwidth and the optimal performance at a single frequency appears in any practical circuit design.  The physics of dc SQUIDs also require trade offs in a good design.  Although modifications to this circuit will change the particulars, the considerations outlined below appear in any superconducting microwave network based on dynamically-modulated SQUIDs~\cite{Duzer,Cast08,Murc12,Mutu13}.        


We first address the schematic-level trade-off.  One must assume that $r$, the transmission line impedance, is practically constrained to 50 $\Omega$ and that we will target designs in which $\omega_0/2\pi$, the center frequency of operation, is in the usual 4-8 GHz band for quantum superconducting microwave networks.  In optimizing the ratio $\omega_0l/r$, this leaves only $l$ as the flexible parameter.  As described in subsection~\ref{sec:Circuit}c, if network bandwidth is held constant, but this ratio increases, ideal circulator operation will decrease.  For example, finite $\omega_0l/r$ produces back reflections (e.g. $|\mathbb{S}_{11}|^2\neq0$ in Fig.~\ref{fig:Admit_IO_Circ_Graphs}).  While it is possible to include additional, lossless matching networks at each port to reduce back reflections over some frequency range, the Bode-Fano criterion suggests that a nonvanishing $\omega_0l/r$ fundamentally limits excellent network matching over broad bandwidths ~\cite{Fano50,Pozar}.  This further supports the intuition gained from Fig.~\ref{fig:Admit_IO_Circ_Graphs} that without a radical redesign of the network, a larger $l$ decreases the performance achievable over a given bandwidth of operation.  

To roughly match the performance of commercial ferrite circulators, we can allow ourselves nonidealities in the signal routing at the -20 dB level.  We see such performance in Fig.~\ref{fig:Admit_IO_Circ_Graphs}c, in which $l=1$ nH.  Inductances of this scale are achievable in SQUID networks fabricated using optical lithography~\cite{Sauv95,Mate08}.  Although the performance depicted in Fig.~\ref{fig:Admit_IO_Circ_Graphs}b is better than in c, the assumed parameters are not realistic.  The baseline impedance is a reasonable $l=0.5$ nH, but Fig.~\ref{fig:Admit_IO_Circ_Graphs}b assumes $\epsilon=1$, which implies that the SQUIDs periodically achieve infinite inductance (the feasibility of $\epsilon=1/\sqrt{2}$ will be discussed below).  Infinite inductance is obviously unphysical, but the actual limitations require some discussion of SQUID physics.

\begin{figure}
\includegraphics[width=.5\textwidth]{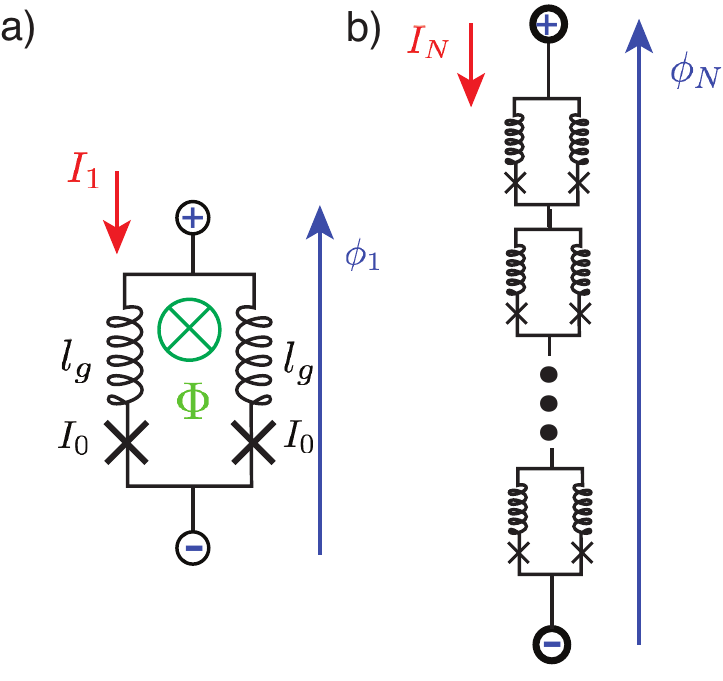}\\
\caption{a) Schematic of a dc SQUID with $I_0$ critical current Josephson junctions, $l_g$ geometric inductance in each branch, and a magnetic flux $\Phi$.  b) A series array of $N$ such SQUIDs.}\label{fig:SQUIDs}
\end{figure}

A single dc SQUID is formed by two Josephson junctions connected in parallel by superconducting wires, forming a loop, Fig.~\ref{fig:SQUIDs}a.  For simplicity, we assume the junctions are identical and junction capacitance is negligible, and we ignore the geometric inductance of the wires for the moment.  Then, Faraday's law and the Josephson relations allow us to relate the voltage across ($V_1$) and current flowing between ($I_1$) two leads connected on opposite sides of a single SQUID loop in the presence of an external magnetic field~\cite{Duzer}.  Assuming null initial conditions, we identify a single SQUID's effective inductance $L_1$ as the ratio between the current and branch flux ($\phi_1(t) = \int_{-\infty}^tV_1(\tau)d\tau$) in the SQUID leads.  SQUID inductance is nonlinear, as this ratio depends on $I_1$, and we find that to third order in $I_1/I_s$
\begin{eqnarray}\label{eq:LSQUID}
\phi_1(t) &=& L_1(t)I_1(t),\text{ with}\\\label{eq:nLSQUID}
L_1(t) &\approx& l_s(t)\left(1+\frac16(I(t)/I_s(t))^2\right),\\
l_s(t) &=& \frac{\varphi_0}{I_s(t)} = \frac{\varphi_0}{2I_0}\left|\sec\left(\frac{\Phi(t)}{2\varphi_0}\right)\right|
\end{eqnarray}
where $\varphi_0 = \hbar/2e$ is the reduced flux quantum, $I_s$ is the SQUID critical current (the maximum current flowing between these two leads that the junction can support in its superconducting state), $I_0$ is the critical current of a single junction, and $\Phi(t)$ is what the magnetic field flux through the loop would be in the absence of screening currents.  Thus, for small currents, a SQUID acts like a linear inductor with a magnetic flux-dependent inductance.  The relative magnitude of the linear and nonlinear inductances, $1/6$, is fixed: a larger inductor requires a SQUID that saturates at smaller currents.  

Now consider an array of $N$ identical SQUIDs with their leads connected in series, Fig.~\ref{fig:SQUIDs}b.  In this case, the current flowing through each, and the voltage across each are identical.  Thus, the relationship between the branch flux across and the current through the entire array is $\phi_N = L_NI_N$, where $L_N = NL_1$.  Qualitatively, adding SQUIDs in series allows us to increase linear inductance without decreasing the saturation current.  Quantitatively, to third order in $I_1/I_a$ we have that~\cite{Cast08,Eich14}
\begin{eqnarray}\label{eq:nLArray}
L_N(t) &\approx& l_a(t)\left(1+\frac{1}{6N^2}(I(t)/I_a(t))^2\right),\\
l_a(t) &=& \frac{\varphi_0}{I_a(t)}, \quad I_a(t) = I_s(t)/N,
\end{eqnarray}      
from which we see that for a fixed linear array inductance $l_a$, the nonlinear inductance scales as $N^{-2}$.  Less nonlinearity means higher saturation powers in analog signal processing. 

The SQUIDs' nonlinearity limits the circulator's power handling.  As the SQUID inductance increases with current, center frequencies of microwave resonators containing SQUIDs typically decrease as the power of incident signals increase.  The approximation of a SQUID as a linear inductor tends to break down when this center frequency shift is comparable to the resonator bandwidth.  In the case of high quality resonators, it is convenient to reparameterize the nonlinear inductance in Eqs.~\eqref{eq:nLSQUID} and~\eqref{eq:nLArray} as an effective ``Kerr constant'' that gives this center frequency shift per microwave photon stored in the resonator (i.e. stored microwave energy in units of $\hbar\omega_0$)~\cite{Cast08,Eich14}.  Up to a factor of order unity (depending on the resonator construction) this Kerr constant is
\begin{equation}\label{eq:K}
K \approx -\frac{\hbar\omega_0^2}{I_s^2l_a}\propto N^{-2}, 
\end{equation} 
where the sign of $K$ indicates the direction of the center frequency shift.  For example, using parameters compatible with Fig.~\ref{fig:Admit_IO_Circ_Graphs}c, $\omega_0 = 2\pi\times 6.16$ GHz, $l_a=1$ nH, and assuming an array of $N=20$ SQUIDs compatible with junctions produced by optical lithography in the NIST NbAlOxNb trilayer process~\cite{Sauv95,Mate08}, $I_s=6.6\,\mu$A, one finds $K\approx -2\pi\times580$ kHz.  With this Kerr constant, and assuming the resonator bandwidth depicted in Fig.~\ref{fig:Admit_IO_Circ_Graphs}c, one would expect the nonlinear reactance of the SQUIDs to become noticeable with of order 100 photons stored in the resonator (because 241 MHz = $(2\pi)^{-1}|K|\times415\text{ photons}$).  A comparable design based on single SQUIDs would require $I_s=0.33\,\mu$A and would exhibit nonlinearities well below 10 stored photons.  For quantum superconducting microwave experiments, the saturation power of commercial ferrite circulators is effectively infinite.  While experiments involving one or two superconducting qubits often operate at signal powers that do not saturate routers and amplifiers networks with single-SQUID-scale saturation currents, higher saturations powers are more critical in quantum networks that employ more weakly coupled systems (e.g. mechanical oscillators and magnetic spins), applications outside of quantum information (e.g. astrophysical detectors), and in current and future quantum networks with more signal multiplexing.


Optimizing our circuit design according to the prescriptions thus far, one might conclude that the best strategy is to employ SQUID arrays with $l_a\approx1$ nH (with no external flux bias) as our tunable inductors, and let $N\rightarrow\infty$, $I_0\rightarrow0$ to minimize nonlinear effects.  This is imprudent for several reason.  One is a larger circuit footprint, which adds stray geometrical inductance throughout the circuit.  Other reasons are due to the geometric inductance of the SQUID loop wires themselves, $l_g$ in Fig.~\ref{fig:SQUIDs}a, which we've ignored thus far.  This SQUID loop inductance limits both the minimum $l_{s,\text{min}}$ and maximum $l_{s,\text{max}}$ achievable through flux tuning of a SQUID array.  Both constraints worsen as $\varphi_0/I_0l_g$ decreases~\cite{Duzer}.  With minimum and maximum array inductances in general one finds that   
\begin{equation}
|\epsilon|<\frac{\eta^2-1}{\eta^2+1},\quad\eta=l_{s,\text{max}}/l_{s,\text{min}}.
\end{equation}
In practice, $\eta$ seems to be limited to about 4 in superconducting microwave circuits with of order 10-100 SQUIDs in an array~\cite{Cast08}, which limits $|\epsilon|<0.88$.  Even longer arrays would be even less tunable.  Constraints on the maximum $\epsilon$ ultimately limit circulator bandwidth, which is proportional to $\epsilon^2$ (section~\ref{sec:Circuit}B).  The simulation depicted in Fig.~\ref{fig:Admit_IO_Circ_Graphs}c, with $\epsilon=1/\sqrt{2}$, is consistent with this tunability constraint, and achieves a bandwidth of 241 MHz.  Commercial circulators used in contemporary superconducting microwave experiments have bandwidth of a few GHz, but the vast majority of this bandwidth is unused.  Most circuits involved in routing quantum microwave signals in contemporary experiments operate at bandwidths of order 10 MHz or smaller~\cite{Mall11,Murc12,Narl14,Abdo14,Cast08,Mutu13}.  Moreover, the center frequency of our network may be tuned broadly by varying the uniform background field, as depicted in Fig.~\ref{fig:BSchem}.  Thus, while not as broadband as commercial options, this network should have an achievable bandwidth more than sufficient for most applications in superconducting microwave networks. 

\subsubsection{Input-output response with SQUID-like inductance modulation}

\begin{figure*}
\includegraphics[width=1\textwidth]{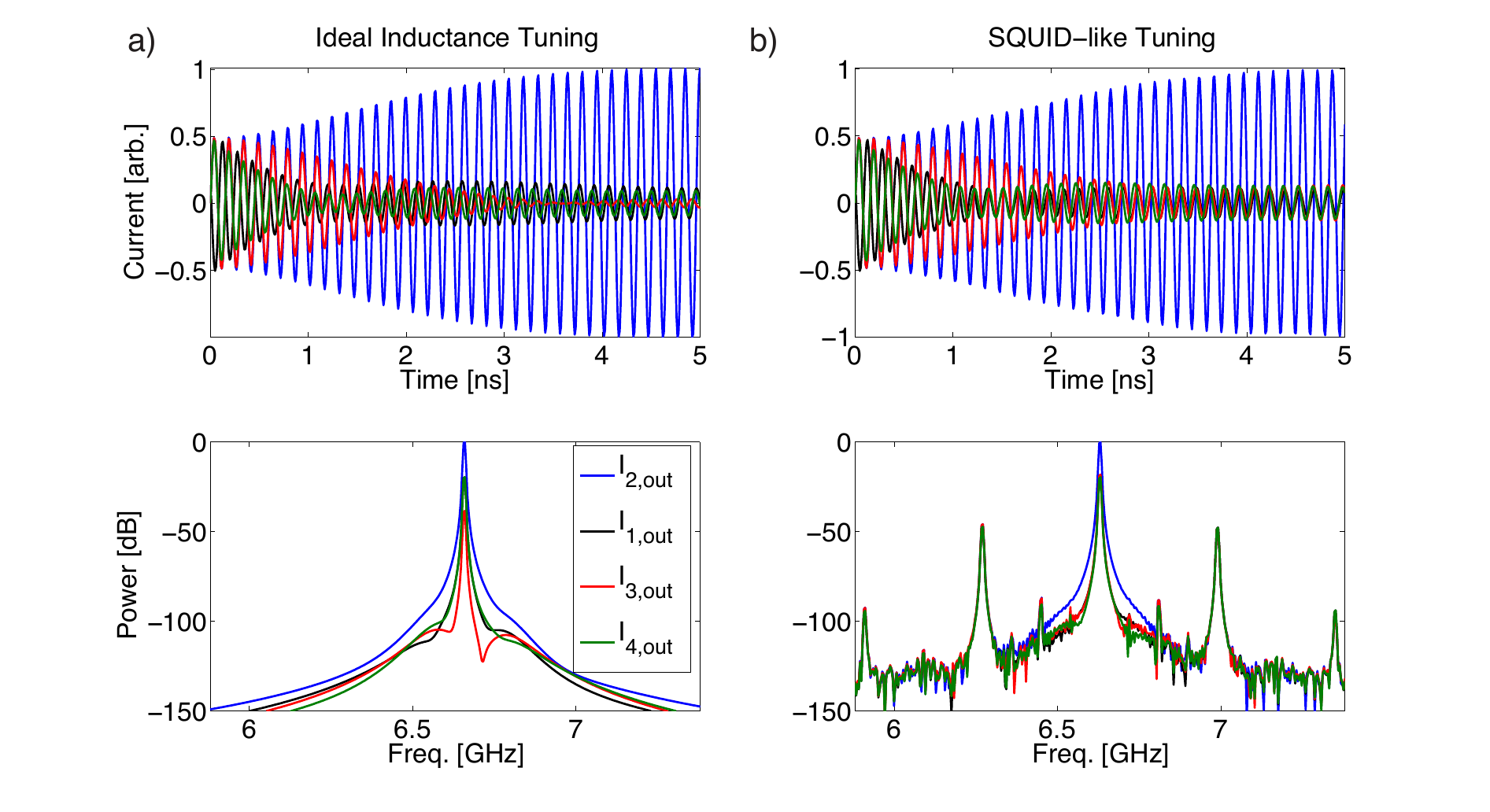}\\
\caption{Numerical network simulations.  a) Network output response in time and frequency, driving port 1, given parameters from Fig.~\ref{fig:Admit_IO_Circ_Graphs}c, and the ideal inductance modulation given in Fig.~\ref{fig:TCoupling}d and Eq.~\eqref{eq:LsIdeal}.  b) The same network simulation, but where the linear inductances are modulated in a ``SQUID-like'' manner, i.e. using Eq.~\eqref{eq:LsSquid}, as described in the text.}\label{fig:ITuning}
\end{figure*}

We saw in section~\ref{sec:Circuit} that proper periodic modulation of the inductances according to Fig.~\ref{fig:TCoupling}d causes the network to scatter signals without creating any sideband excitations.  Using a numerical, time-domain, MATLAB Simscape simulation of the circuit depicted in Fig.~\ref{fig:TCoupling}d, network performance in this ideal case is depicted in Fig.~\ref{fig:ITuning}a.  Using the same parameters as in Fig.~\ref{fig:Admit_IO_Circ_Graphs}c, $r=50\,\Omega$, $l=1$ nH, $c=1$ pF, $\epsilon=1/\sqrt{2}$, $\Omega = \Omega_0=2\pi\times99$ MHz, a continuous-wave, input current drive at fequency $\omega_d=2\pi\times6.66$ GHz is applied to port 1 suddenly at time $t$=0.  The top plot in Fig.~\ref{fig:ITuning}a depicts the resulting current measured in the output signals for the first 5 ns of the simulation (after separating input from output signals in the simulation).  Initially, the output current is equally distributed among the four ports, but after a few ns nearly all of the output current is measured at port 2.  The bottom plot gives the normalized output current power spectrum for a 250 ns simulation (i.e. well into steady state).  The spectra are sharply peaked at $\omega_d$, with the peak of the $I_{1,\text{out}}$, $I_{3,\text{out}}$, and $I_{4,\text{out}}$ signals at least 20 dB below that of $I_{2,\text{out}}$.  While the shoulders are slightly fat, there are no visible sidebands.  Driving the inputs of the other ports produces responses identical to Fig.~\ref{fig:ITuning}a, up to a cyclical permutation of the port indices.

Unfortunately, it is not simple to tune a SQUID's inductance exactly as 
\begin{eqnarray}\label{eq:LsIdeal}
l_s^\prime(t)&=&l(1\pm\epsilon\cos(\Omega t))^{-1}\text{ or }l_s^\prime(t) = l(1\pm\epsilon\sin(\Omega t))^{-1}, 
\end{eqnarray}
as in Fig.~ \ref{fig:TCoupling}d.  Because the linear inductance of a SQUID or array of SQUIDs goes as Eq.~\eqref{eq:Ls}, applying a simple background magnetic flux that consists of a static and sinusoidal component yields effective inductances that vary as
\begin{eqnarray}\label{eq:LsSquid}
l_s(t) &=& \frac{\varphi_0}{2I_0\left|\cos\left(\frac{F(t)}{2\varphi_0}\right)\right|},\nonumber\\
\Phi(t) &=& \Phi_{\Sigma}\pm \Phi_\Delta\cos(\Omega t)\text{ or }\Phi(t) = \Phi_{\Sigma}\pm \Phi_\Delta\sin(\Omega t).
\end{eqnarray} 
It is possible to make $l_s$ exactly equal to $l_s^\prime$ in the limit $\varphi_0/2I_0\rightarrow0$, $\Phi_\Sigma/2\varphi_0\rightarrow\pi/2$.  However, this limit corresponds to an infinitely flux tunable Josephson inductance, which is unphysical.  Instead, it is more realistic to set $\varphi_0/2I_0=2l$, where $l$ is the desired $l$ in Eq.~\eqref{eq:LsIdeal}, and set $\Phi_\Sigma/2\varphi_0=\pi/3$ so that $l_s=l_s^\prime$ when $\Phi_\Delta=0$.  One then chooses $\Phi_\Delta$ such $l_s\approx l_s^\prime$, given the desired $\epsilon$ in Eq.~\eqref{eq:LsIdeal}.  

We can now redo the simulation in Fig.~\ref{fig:ITuning}a, but with a variable inductance model based on Eq.~\eqref{eq:LsSquid} rather than the ideal Eq.~\eqref{eq:LsIdeal}.  Choosing parameter values that closely emulate the simulation in Fig.~\ref{fig:ITuning}a, $r=50\,\Omega$, $\varphi_0/2I_0=2$ nH, $\Phi_\Sigma/2\varphi_0=\pi/3$, $c=1$ pF, $\Phi_\Delta/2\varphi_0=0.38$, $\Omega = 2\pi\times90$ MHz, $\omega_d=2\pi\times6.63$ GHz (parameters found through trial-and-error optimization), produces the analogous plots in Fig.~\ref{fig:ITuning}b.  The time-domain response is nearly identical to the ideal case.  The most noticeable difference is that the $I_{3,\text{out}}$ response is about the same as $I_{1,\text{out}}$ and $I_{4,\text{out}}$.  The differences between the steady state power spectra are more obvious.  The output signals now exhibit sidebands, which are caused by the slight differences between Eqs.~\eqref{eq:LsSquid} and~\eqref{eq:LsIdeal} for these parameters.  Happily, though, the largest of these sidebands are nearly -50 dB$_c$ and are the 3rd harmonics of $\Omega$ (i.e. are at $\omega_d\pm4\Omega$).  The lower harmonics of $\Omega$ are almost completely suppressed by the symmetry of the circuit, which is remarkable considering the significant, $\Phi_\Delta/2\varphi_0=0.38$, modulations of the SQUID-like inductances.   

A quantitative analysis of the spectrum observed in Fig.~\ref{fig:ITuning}b is beyond the scope of this article, but will be considered in a companion publication (which will also contain a quantized model of the circulator)~\cite{Lalu15}.  This is done by first deriving an Hamiltonian describing the circulator including its input and output ports. In a second step, Yurke's approach to input-output formalism is used to compute the scattering properties of the circuit~\cite{Yurk84}. In contrast to the analysis in section \ref{sec:time}, following Yurke allows us to obtain exact, time-domain equations of motions for the circuit's degrees of freedom.  Using theses results, the spectrum observed in Fig.~\ref{fig:ITuning}b can be understood quantitatively.

\section{Conclusion}

We have introduced and analyzed a novel design for a four-port circulator based on time-varying inductances (Fig.~\ref{fig:TCoupling}d) that has the potential to replace the lossy and non-integrable commercial ferrite circulators ubiquitous in superconducting quantum microwave experiments.  The basic module is a dynamically-unbalanced bridge network of four SQUIDs (or SQUID arrays) \cite{Berg10}, depicted in Fig.~\ref{fig:BSchem}.  These bridges dynamically modulate the coupling between four transmission lines and two resonant modes, and the construction ensures that this coupling is the only aspect of the dynamics that is modulated.  The construction also ensures that when the inductances are tuned perfectly, the network's output signals are not complicated by frequency, phase, or amplitude modulation.  This critical feature is greatly aided by limiting the dynamic modulation to the coupling between the resonant modes and the transmission lines.  As a result, the resonant modes act as if they are ``spinning'' relative to the transmission lines or, equivalently, as if they are coupled by a synthetic magnetic field, Fig.~\ref{fig:TCirc}.   

The design has a number of attractive features such as modulation pump tones detuned from the signal by more than a decade, a broadly tunable center frequency, a tunable bandwidth, high saturation power, and a lumped element and modular construction.  We have given an overview of many of the considerations needed to realize a design capable of achieving a 240 MHz bandwidth and a saturation power of order 100 microwave photons per inverse bandwidth with superconducting microwave technology.  The design of a practical electromagnetic circulator is a difficult and important problem and different technological constraints have yielded a wide variety of solutions and potential solutions over the decades, of which Refs. \cite{Fay65,Alle56,Auld59,Ande65,Ande66,Este14,Tana65,Kama11,Fang12,Ranz14,Abdo14,Koch10,Viol14,Tell48} are but a small sample.  Because of this long tradition, we analyzed this network from several different perspectives, through analogy to familiar, ferrite-based circulator designs, through lumped element circuit analysis, through approximate input-output models (appropriate for distributed RF or optical systems), in the time and frequency domain, analytically, and numerically.  We expect that the present design will find near-term use context of quantum superconducting microwave networks, but we also hope that this diverse analysis will help inspire an even greater diversity of future solutions and perspectives in even broader contexts.  

This work is supported by the ARO under the contract W911NF-14-1-0079.  K. Lalumi\`{e}re and A. Blais acknowledge support from the ARO under grant W911NF-14-1-0078, as well as from NSERC and CIFAR.  J. Kerckhoff would like to thank John Gough for a useful discussion.

\begin{appendix}

\end{appendix}

\end{document}